\documentclass{aa}

\usepackage{graphicx}
\usepackage[colorlinks=true, linkcolor=blue, citecolor=blue, urlcolor=blue]{hyperref}

\usepackage{txfonts}
\usepackage{xcolor}
\usepackage{comment}
\usepackage{multirow}

\defcitealias{li22}{LL22}
\usepackage[switch]{lineno}
\usepackage{placeins}
\usepackage{float}

\begin{document}

   \title{Binary black holes in magnetized AGN disks}

   \author{Raj Kishor Joshi
          \inst{1,2}\and
          Aryan Bhake\inst{2}\and
          Biswajit Banerjee\inst{3} \and
          Bhargav Vaidya\inst{2}\and                     
         Milton Ruiz\inst{4}\and
         Antonios Tsokaros\inst{5,6,7}\and
         Andrea Mignone\inst{8}\and
        Marica Branchesi\inst{3}\and
        Amit Shukla\inst{2}\and
        Miljenko \v{C}emelji\'{c}\inst{9,1,10}
        }

   \institute{Nicolaus Copernicus Astronomical Center of the Polish Academy of Sciences, Bartycka 18, PL-00-716 Warsaw, Poland \\
              \email{rjoshi@camk.edu.pl}
         \and Department of Astronomy, Astrophysics and Space Engineering, Indian Institute of Technology Indore, Indore 453552, India
         \and Gran Sasso Science Institute (GSSI), I-67100 L’Aquila, Italy
         \and Departament d’Astronomia i Astrof\'{i}sica, Universitat de Val\`{e}ncia, C/ Dr Moliner 50, 46100, Burjassot (Val\`{e}ncia), Spain
         \and Department of Physics, University of Illinois at Urbana-Champaign, Urbana, Illinois 61801, USA
        \and National Center for Supercomputing Applications, University of Illinois at Urbana-Champaign, Urbana, Illinois 61801, USA
        \and Research Center for Astronomy and Applied Mathematics, Academy of Athens, Athens 11527, Greece
        \and Dipartimento di Fisica, Universit\'a degli Studi di Torino, Via Pietro Giuria 1, I-10125 Torino, Italy
        \and Nicolaus Copernicus Superior School, College of Astronomy and Natural Sciences, Gregorkiewicza 3, 87-100, Toru\'{n}, Poland
        \and Research Centre for Computational Physics and Data Processing, Institute of Physics, Silesian University in Opava, Bezru\v{c}ovo n\'am.~13, CZ-746\,01 Opava, Czech Republic
             }

  \abstract
   {Stellar-mass binary black hole (BBH) mergers occurring within the disks of active galactic nuclei (AGN) are promising sources for gravitational waves detectable by the LIGO, Virgo, and KAGRA (LVK) interferometers. Some of these events have also been potentially associated with transient electromagnetic flares, indicating that BBH mergers in dense environments may be promising sources of multi-messenger signals. To investigate the prospects for electromagnetic emission from these systems, we study the dynamics of accretion flows onto BBHs embedded in AGN disks using numerical simulations. Although recent studies have explored this scenario, they often employ simplified disk models that neglect magnetic fields. In this work, we examine how strong magnetic fields influence and regulate the accretion onto such binary systems. In this context, we conduct three-dimensional magnetohydrodynamical local shearing-box simulations of a binary black hole system embedded within a magnetized disk of an AGN. 
   We observe that the dynamically important magnetic fields can drive the formation of well-collimated outflows capable of penetrating the vertical extent of the AGN disk. However, outflow generation is not ubiquitous and strongly depends on the radial distance of the binary from the supermassive black hole (SMBH).  In particular, binaries placed at a larger distance from the central SMBH show relatively more transient accretion and the formation of stronger spiral shocks. Furthermore, accretion behavior onto the binary system via individual circum-singular disks (CSDs) is also modulated by local AGN disk properties. Our simulations highlight the importance of shear velocity in the amplification of the toroidal magnetic field component, which plays a crucial role in governing the outflow strength.  
   }


   \maketitle
%
\section{Introduction}

Gravitational wave (GW) detections of merging binary stellar-mass black holes (BBHs), accumulated over several observational runs (O1–O4) by the LIGO, Virgo, and KAGRA (LVK) collaborations \citep{abbott20,2023abbott, LIGOScientific:2016aoc, LIGOScientific:2018mvr, LIGOScientific:2020ibl, KAGRA:2021vkt}, have sparked extensive theoretical efforts to understand how black hole binaries form \citep[][and references therein]{Sedda:2021vjh, Mapelli:2020vfa, Mapelli:2021taw, Bouffanais:2021wcr, Mandel:2021smh, Grobner:2020drr, 10.1111/j.1365-2966.2012.21486.x}. Currently, the astrophysical environments in which such BBHs form are not well understood, and several channels have been proposed as plausible origins for the formation of these merging systems \citep{gayathri21, Mapelli:2020vfa, Mapelli:2021taw, Ford:2021kcw, Mandel:2021smh, Mestichelli:2024djn, 2021ApJ...913L...7A, Yang:2019cbr}. 

GW signals from these events enable us to infer the distribution of mass, spin, and possible eccentricities of BBHs, which is crucial information to unveil their origin and evolution. However, GW observations alone provide limited insight into the surrounding environment. This limitation can be addressed through the detection of potentially associated electromagnetic (EM) counterparts, which offer complementary environmental information.

In contrast to binary neutron star mergers \citep[i.e., GW170817/ GRB 170817A;][]{LIGOScientific:2017vwq, LIGOScientific:2017ync}, production of EM counterparts from BBH mergers is more complex.
EM signals associated with BBH mergers are detectable only when the mergers occur within a gas-rich environment\footnote{The detection of faint hard X-ray signal in Fermi/GBM, GW150914-GBM \citep{Connaughton:2016umz}, possibly associated with GW150914 \citep{LIGOScientific:2016aoc} boosted this line of thought.}\citep[][and references therein]
{Perna:2016jqh, Yamazaki:2016fyr,Li:2021hed, Grobner:2020drr, Rodriguez-Ramirez:2024ikd, Graham:2020gwr, Chen:2023xrm, Bartos:2016dgn}.

The GW event, GW190521, detected on 21 May 2019 at 03:02:29 UTC with a false alarm rate of 3.8$\times10^{-9}$ Hz \citep{LIGOScientific:2020iuh, LIGOScientific:2020ufj} is considered to be an exceptional event because of its GW properties. The merger produced a black hole remnant with a mass of approximately $150 M_{\odot}$, originating from progenitor black holes of about $85 M_{\odot}$ and $66 M_{\odot}$ \citep{LIGOScientific:2020iuh}. These unusually high component masses exceed the typical limits predicted by standard binary black hole formation scenarios through the evolution of stellar remnants \citep{kinugawa14, belczynski16, LIGOScientific:2020ufj,Woosley:2016hmi, Rodriguez:2019huv, Fragione:2020han}. 
Although there are attempts to explain the formation of this type of system through isolated binary evolution \citep{Cui:2023tfo}, the hierarchical merger channel more naturally explains the origin of such massive binary components. Such mergers are expected to occur in dense nuclear star clusters (NSCs) because of many-body encounters \citep{mouri2002,hopman06,rodriguez16,fujii17,McKernan18,wh24}. In the local universe, the highest density of stellar-mass black holes appears to be in the Galactic center, as suggested by observational evidence \citep{generozov18, hailey18} and earlier theoretical predictions \citep{morris93,miralda2000}. {Additionally, the dense gaseous accretion disks orbiting the SMBH also provide a favorable environment for BBH formation and mergers. The efficiency of the gas-assisted binary formation mechanism in the context of AGN channels has been recently investigated through analytical works \citep{fm22,vmp24} and simulations \citep{thk20,tkh21,rwk24,wrk25}. These studies indicate that the substantially shorter average lifetimes of binaries in AGN disks lead to a higher probability of merger, making BBH mergers in AGN disks a non-negligible contributor to the BBH merger rate observed by the LVK. 
A fraction of black holes naturally have orbits that are embedded within the AGN disk, while others can gradually migrate into the disk due to interactions with the gas \citep{artymowicz93,goodman04}. Once embedded in the disk, these black holes can undergo dynamical interactions and accretion-driven migration, enhancing the likelihood of binary formation and merger \citep{lgmf18, deh23, rwk24,wh24, ddl25, wlw25}.}

The LVK collaboration localized the GW source in an area of approximately 765 $deg^{2}$. Approximately 34 days later, the Zwicky Transient Facility (ZTF) identified a flaring event from an AGN J1249+3449 \citep{Graham:2020gwr}. ZTF surveyed nearly half of the localization region with 47 $deg^{2}$ field of view and reported the flaring event. The AGN is located at a redshift of $z=0.438$ \citep{Grobner:2020drr}. Additionally, BBHs embedded in AGN disks are expected to be strongly lensed by the central supermassive black hole. The detection of lensed GW signals could thus serve as a unique probe to determine whether AGNs represent a significant formation channel for BBHs \citep{Leong_2025}.
This has led to a growing interest in multi-messenger (MM) astrophysics from BBH mergers, where the joint analysis of different observational channels enhances our understanding of the physical conditions, formation mechanisms, and aftermath of compact object mergers. 

Once a bound BBH forms in an AGN disk, its evolution depends on the properties of the surrounding gas, in addition to intrinsic binary parameters such as separation, orbital frequency, and mass ratio of the components \citep[e.g.][]{Tsokaros:2022hjk}.
One natural approach to address the problem of accretion onto BBH is to treat this as a circumbinary accretion problem \citep{noble2012,armagol21}, where the AGN disk supplies gas to a {circumbinary disk} (CBD) around the binary. \cite{li22}(hereafter \citetalias{li22}) have shown that the strong velocity shear and angular momentum can significantly suppress inflow to binary systems; therefore, circumbinary accretion models may not directly translate to binaries embedded in large-scale AGN disks \citep{lm23}. 
Recent numerical simulations have extensively explored the interaction between embedded BBHs and their surrounding disks, primarily through detailed two-dimensional (2D) simulations \citep[\citetalias{li22}]{Baruteau2011,Li2021,Li2022,lilai23,lilai24}. A smaller number of studies has extended this analysis to three dimensions (3D) \citep{Dempsey_2022,wh24,Dittmann_2024,Calcino_2024}, offering additional insights into the complex dynamics at play. One crucial missing ingredient in these simulations is the background magnetic field of the AGN disk. There is a handful of simulations that have incorporated the effect of the magnetic field \citep{mc24, rcrm25,most24}. 
\cite{most24} carried out high-resolution, 3D Newtonian magnetohydrodynamics (MHD) simulations to explore the possibility of magnetically arrested accretion flows for BBH surrounded by a CBD. Their findings reveal that circumbinary accretion flows also achieve a magnetically arrested state similar to single accreting
black holes, and the accretion onto the binary system occurs in periodic bursts driven by magnetic flux eruptions \citep{nia03,tnm11,dvf22}. This behavior is a significant aspect to consider when investigating potential EM counterparts. Building on jet breakout models, \cite{tagawa23} demonstrated that jets originating from accreting BBH mergers within AGN disks may manifest as distinctive transients observable in infrared, optical, and X-ray wavelengths.\\

{The dynamics of the black hole–gas interaction on the horizon scale has also been addressed through full general relativistic numerical simulations. The studies presented in \cite{Gold:2013zma,Gold:2014dta} examined the dependence of GW and EM signals on the binary mass ratio and suggested that transient variations in jet luminosity might provide clues for distinguishing mergers of black holes in AGNs from single accreting black holes based solely on jet morphology.
Numerical simulations reported in \cite{giaco22, kprs18, armagol21, rts23} suggest a consistent formation of jets launched from both black holes, regardless of the binary mass ratio, provided the surrounding plasma remains in a force-free regime.
These jets tend to align with the spin axes of the black holes \citep{rts23}. The results from simulations reported by \cite{fcgc24} highlight the critical role of the surrounding environment and magnetic fields in shaping the observable EM counterparts of black hole mergers. This dependence arises because the structure and composition of the ambient medium influence accretion dynamics and jet propagation \citep{cg24}.}\\
 
Following the insights obtained from these results, we conduct MHD simulations of a BBH system embedded within a magnetized AGN disk. Unlike previous studies, we do not start our simulations with BBH surrounded by a circumbinary disk \citep{munoz20,duffell24}, which supplies gas to the binary. Instead, in our simulations, the BBH system is embedded inside a local patch of the AGN disk, popularly known as the `shearing-box' model. This patch rotates around the SMBH in a circular orbit, and the gravitational force arising from the SMBH is also taken into account. The central aim of this study is to evaluate the influence of magnetic fields on the development of outflows {and to investigate how local disk shear regulates accretion and outflow formation.}
\\

The structure of this paper is as follows: Section \ref{sec:meth} outlines the governing equations, numerical setup, and initial and boundary conditions used in our simulations. Section \ref{sec:res} presents the results of these simulations. In Section \ref{sec:con}, we summarize our work.

\section{Methods}
\label{sec:meth}
\begin{figure}
   \centering
   \includegraphics[width=\columnwidth]{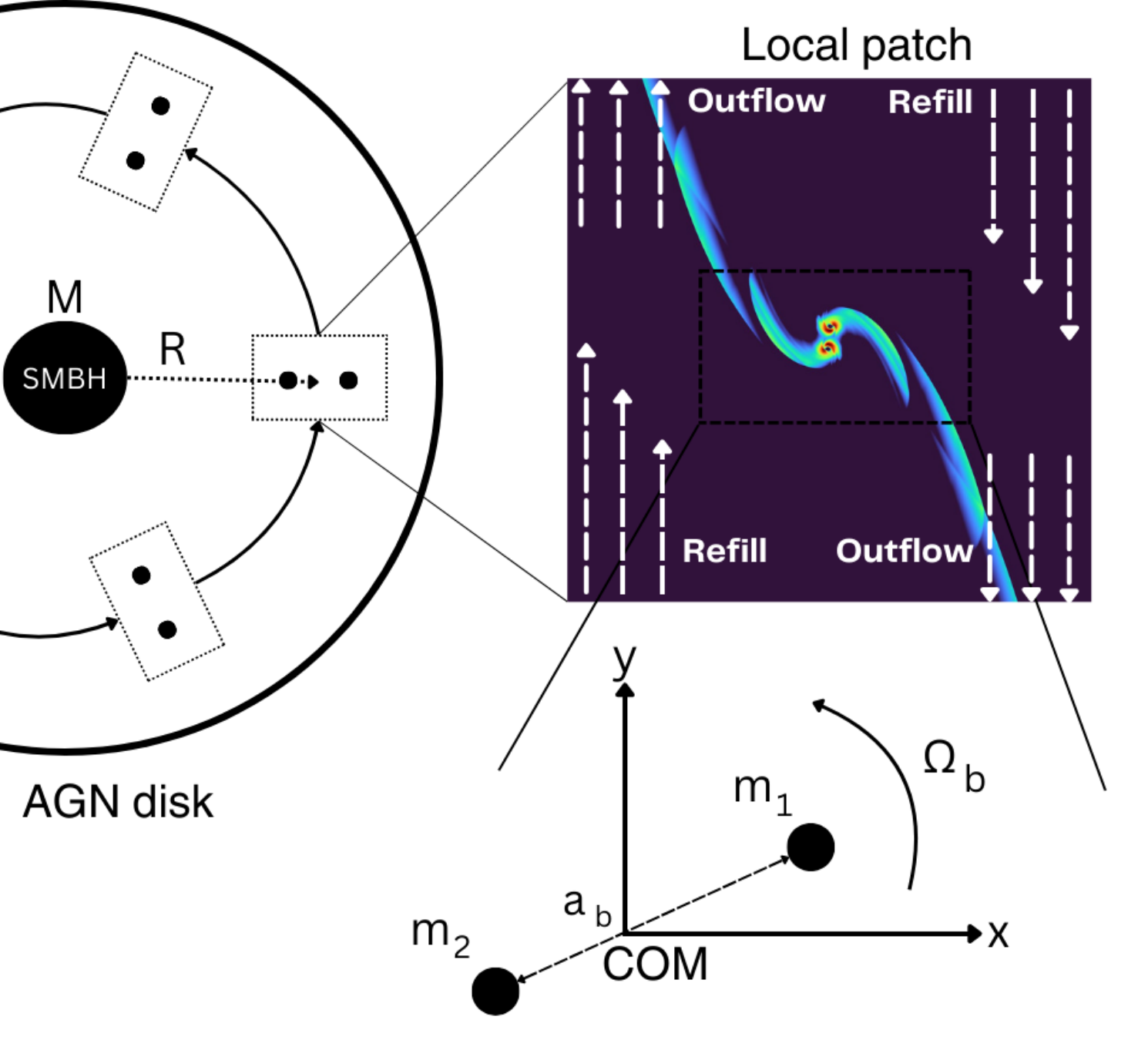}
      \caption{Cartoon depiction of our model and simulation setup to study the evolution of BBH ($m_\text{1}$ and $m_\text{2}$) with a separation $a_\text{b}$, embedded in the disk of a SMBH of mass $M$. CoM of binary moves in a circular orbit in the $x-y$ plane.}
         \label{fig:schem}
   \end{figure}

We simulate the equal-mass BBH embedded in the Keplerian disk of an AGN using the publicly available code \texttt{PLUTO} \citep{pluto07}. In this three-body system, two different length scales are involved: the first one is the distance of BBH's center of mass (CoM) from the AGN, and the second is the separation of BBH components. Due to the difference between these two scales, resolving the features of accretion flows around individual components is computationally expensive in global disk simulations. To circumvent this problem, we use a localized co-rotating box of the accretion disk popularly known as the ``shearing box'' model \citep{Goldreich_1965,hgb95}. With this approximation, we transform the global cylindrical geometry of the disk into local Cartesian coordinates, with unit vectors \(\hat{x}\), \(\hat{y}\), and \(\hat{z}\). The CoM of the BBH is located at the position \((x, y, z) = (0, 0, 0)\) and corresponds to a reference radius \(R\) from the SMBH as shown in Fig. \ref{fig:schem}. The Keplerian velocity is \(V_\text{K} = \sqrt{GM/R}\), \ M \ is the mass of the SMBH,  and the Keplerian frequency is \(\Omega_\text{K} = V_\text{K} / R\). This is the frequency of rotation of our reference frame. Similar to previous studies \citep[\citetalias{li22}]{most24, Baruteau2011},  we do not resolve the horizons or capture the strong-field relativistic effects near the individual black holes and adopt a Newtonian framework. Within this approximation, we solve the equations of ideal MHD

\begin{equation}
\frac{\partial \rho}{\partial t}+\nabla \cdot\left(\rho \mathbf{v}\right)=0\, ,  
\label{eq:continuity}
\end{equation}

\begin{equation}
\frac{\partial (\rho \mathbf{v})}{\partial t}+\nabla \cdot\left(\rho \mathbf{v\otimes  v-B\otimes B}\right)+\nabla p_\text{t}=\rho\mathbf{g}_\text{s}-2\rho\Omega\hat{z}\times \mathbf{v}+\rho\mathbf{g}_\text{b}\,  ,  
\label{eq:euler}
\end{equation}

\begin{equation}
\frac{\partial \mathbf{B} }{\partial t}+\nabla \times\left(\mathbf{v}\times \mathbf{B}\right)=0 \, ,
\label{eq:induction}
\end{equation}

\begin{equation}
\frac{\partial E}{\partial t}+\nabla \cdot\left[(E+p_\text{t})\mathbf{v}-(\mathbf{v\cdot B})\mathbf{B}\right]=\rho\mathbf{v}\cdot\mathbf{g}_\text{s}+\rho\mathbf{v}\cdot\mathbf{g}_\text{b} \, ,
\label{eq:energy}
\end{equation}
where $\mathbf{g}_\text{s}$ is the tidal expansion of the effective gravity of the central SMBH, given as $\mathbf{g}_\text{s}=\Omega_{\text{K}}^2(2q_\text{sh}x\hat{x}-z\hat{z})$. The second term on RHS of Eq. (\ref{eq:euler}) is the Coriolis force, and $q_\text{sh}$ is the shear parameter. $p_\text{t} = p + p_\text{mag} $ is the
total pressure accounting for thermal ($p$), and magnetic ($p_\text{mag}=B^2/2$)
pressure. The total energy density $E$ is the sum of internal, kinetic and magnetic energy density 
\begin{equation}
E=\frac{p}{\gamma-1}+\frac{1}{2}\rho|\mathbf{v}|^2+\frac{|\mathbf{B}|^2}{2} \, ,
\label{tot_e}    
\end{equation}
while $\mathbf{g}_\text{b}$ is the effective gravitational acceleration due to BBH gravity, given as 

\begin{equation}
\mathbf{g}_\text{b}(\mathbf{r},t)=-\frac{Gm_\text{1}|\mathbf{r}-\mathbf{a}_1(t)|\hat{d_\text{1}}}{((\mathbf{a_\text{1}}(t)-\mathbf{r})^2+\xi^2)^{3/2}}-\frac{Gm_\text{2}|\mathbf{r}-\mathbf{a}_2(t)|\hat{d_\text{2}}}{((\mathbf{a_\text{2}}(t)-\mathbf{r})^2+\xi^2)^{3/2}}\,.
\label{eq:effective_a}    
\end{equation}
In Eq. (\ref{eq:effective_a}), $\mathbf{a}_\text{1}(t)$ and $\mathbf{a}_\text{2}(t)$ represent the time-dependent position vectors of the black holes, while $\mathbf{r}$ denotes the position vector of the grid cell. The unit vectors $\hat{d}_\text{1}$ and $\hat{d}_\text{2}$ point from the black holes toward the grid cell. Furthermore, $\xi=0.01a_{\rm b}$ is the gravitational softening length, $a_{\text{b}}$ is the binary separation. 
In this work, we adopt an ideal gas equation of state with an adiabatic index $\gamma=1.6$, which slightly deviates from the conventional value of $\gamma=5/3$ typically used for a non-relativistic gas. This choice is made to maintain consistency with the simulations reported by \citetalias{li22}, allowing qualitative comparison with their results and facilitating the validation of our methods within a hydrodynamical framework. The binary has a total mass \(m_\text{b} = m_1 + m_2\) and the mean orbital frequency \(\Omega_\text{b}\) given as 

\begin{equation}
\Omega_\text{b} = \sqrt{\frac{G m_\text{b}}{a_\text{b}^3}}\, .
\end{equation} 

We use Harten, Lax, and Van Leer (HLL) Riemann solver \citep{Harten_1983} with piecewise parabolic reconstruction \citep{cw84} and second order Runge-Kutta time stepping method (RK2). The magnetic field is kept divergence-free ($\nabla\cdot\mathbf{B}=0$) utilizing the hyperbolic/parabolic divergence cleaning method \citep{dedner02}. 
We adopt a unit system in which the code units are set to the natural units of the binary system, i.e. \(a_\text{b}\) is the unit of length and the unit of time is taken as \(\Omega^{-1}_\text{b}\).

\subsection{Initial and boundary conditions}

We set up a vertically stratified density profile given by the expression
\begin{equation}
\rho=\rho_0exp\left(-\frac{z^2}{2H_\text{g}^2}\right)\, .
\label{eq:start}    
\end{equation}
In Eq. (\ref{eq:start}), $H_\text{g}=\frac{\sqrt{2c^2_\text{s}}}{\Omega_\text{K}}$ is the pressure scale height of the disk and $c_\text{s}$ is the sound speed. $\rho_0$ is the density in the mid-plane ($z=0$) of the box, we take $\rho_0=1$. {The correspondence between code units and physical units, together with their physical interpretation, is detailed in section \ref{sec:scaling}.}

The following dimensionless parameters describe the simulation environment:

1. Mass ratio of the binary to the central SMBH :
   \begin{equation}
   q_\text{M} = \frac{m_\text{b}}{M}\, .
   \label{eq:param1}  
   \end{equation}

2. Disk aspect ratio \(h\) at disk radius \(R\):
   \begin{equation}
   h = \frac{H_\text{g}}{R} = \frac{c_\text{s}}{V_\text{K}}\,.
   \label{eq:param2}
   \end{equation}
   
3. The ratio of the binary Hill radius \(R_\text{H} \approx R (m_\text{b} / M)^{1/3}\) to \(a_\text{b}\):
   \begin{equation}
   \lambda = \frac{R_\text{H}}{a_\text{b}} = \frac{R}{a_\text{b}} \left( \frac{m_\text{b}}{M} \right)^{1/3}.
   \label{eq:param3}
   \end{equation}

The time-independent background wind profile in the shearing box can be written as

\begin{equation}
    \mathbf{V}_\text{w} 
    = -q_\text{sh} \Omega_\text{K} x \hat{\mathbf{y}}\,. 
\label{eq:back_wind}
\end{equation}

In Eq. (\ref{eq:back_wind}), $q_\text{sh}$ is the local measure of differential rotation \citep{mignone12}, calculated as 
\begin{equation}
q_\text{sh}=-\frac{1}{2}\frac{d\,\log\,\Omega^2(R)}{d\,\log\,R}\,.
\label{eq:q_shear}    
\end{equation}
For the Keplerian disk Eq. (\ref{eq:q_shear}) results in $q_\text{sh}=3/2$.

From the binary's point of view, the flow dynamics is governed by the following characteristic velocity ratios \citep[\citetalias{li22}]{lilai23}
\begin{equation}
\frac{c_\text{s}}{v_\text{b}} = h q_\text{M}^{-1/3} \lambda^{-1/2},
\label{eq:cs_lambda}
\end{equation}

\begin{equation}
\frac{V_\text{s}}{v_\text{b}} = q_{\text{sh}} \frac{\Omega_\text{K}}{\Omega_\text{b}} = q_{\text{sh}} \lambda^{-3/2},
\label{eq:omega_lambda}
\end{equation}

where \(V_\text{s}\) represents the Keplerian shear magnitude over a distance of \(a_\text{b}\) and $v_\text{b}=\sqrt{\frac{G m_\text{b}}{a_\text{b}}}$.
For our setup we fix \(h = 0.01\) and vary \(\lambda=2.5,5,7.5\) for different simulation runs with \(q_\text{M} = 10^{-6}\). It is chosen to be sufficiently large to ensure that the outer boundaries fully contain the spiral arms and encompass the entire Hill sphere. For all simulations, the central region of the domain [-1:1,-1:1,-1:1] has a uniform resolution of $\Delta x=\Delta y=\Delta z=0.01$, the rest of the domain has a resolution  $\Delta x=\Delta y=\Delta z=0.2$. We also performed simulations at lower ($\Delta x=\Delta y=\Delta z=0.02$) and higher resolution ($\Delta x=\Delta y=\Delta z=0.005$) to assess convergence and consistency; the resolution that we have adopted for the results presented in this study was sufficient to capture the essential features of the system without the increased computational expense required by the highest resolution.
The size of the computational domain differs across runs, and the simulation parameters ($\lambda,\,\Omega_\text{K},\,c_\text{s}$) are summarized in Table \ref{tab:param}. The dependence of $c_{\rm s}$ and $\Omega_{\rm K}$ on $\lambda$ is established via equations  \ref{eq:cs_lambda} and \ref{eq:omega_lambda}. 

In the shearing box module \citep{mignone12}, the default boundary conditions are shearing in the $x$ direction and periodic in the $y$ direction, but can be freely assigned in the vertical direction $z$. However, in our setup (illustrated in the schematic diagram in Fig. \ref{fig:schem}), we have instead implemented outflow boundary conditions in the $x$ direction and a combination of refill and outflow boundary conditions in the $y$ direction. {Refilling entails reassigning ghost cells with the initial physical conditions of the ambient disk material. In the ghost zones, the values of density and velocity are calculated using  Eq. (\ref{eq:start}) and (\ref{eq:back_wind}), and pressure ($p=c^2_\text{s}\rho/\gamma$) is calculated from the information of density and sound speed $c_\text{s}$ (given by Eq. \ref{eq:cs_lambda}). In outflow regions, ghost cells adopt the values at the domain boundary. }

\[
\text{At } y = y_{\min}:
\quad
\begin{cases}
x < 0 & \text{refill} \\
x \geq 0 & \text{outflow}
\end{cases}
\]

\[
\text{At } y = y_{\max}:
\quad
\begin{cases}
x \leq 0 & \text{outflow} \\
x > 0 & \text{refill}
\end{cases}
\]

This configuration enables continuous mass injection into the domain to avoid an excessive drop in density resulting from mass loss due to outflows. The black holes are treated as absorbing spheres with a sink radius $r_\text{s}=0.05a_\text{b}$. {After each timestep, all the cells located within a radius $r_\text{s}$ of each binary component $\mathbf{a}_\text{i}(t)\,(i=1,2)$ are identified based on the criterion $|\mathbf{r}-\mathbf{a}_\text{i}(t)|<r_\text{s}$. For these cells, the velocity field is set to zero, and the density and pressure are also set to a smaller value of $10^{-2}$.}    
The magnetic field is injected into the computational domain with inflowing matter, which has a purely vertical component ($B_\text{z}$) with constant value $B_0$ chosen such that the plasma beta $\beta=p/ p_{\text{mag}} =100$ at the midplane. {The choice of a purely vertical initial magnetic field is an assumption based on the previous magnetized shearing box simulations \citep{sasb16,mc24}. A field configuration without an initial toroidal component helps in order to isolate the effects of amplification and winding driven by the binary–disk interaction.}
The magnetic field injection strategy follows \cite{m24}, which allows the BBH system to complete a few orbits before initiating the injection process. We start the injection of the magnetic field around $t=40$. 
In the vertical direction, we have set outflow boundary conditions, and the magnetic field is purely vertical $\partial B_\text{z}/\partial z=0,\, B_\text{x}=B_\text{y}=0$ \citep{bcmr12}. 
{To ensure the consistency and reliability of the modified boundary conditions used in this study, we regenerated the results of a 2D simulation run in hydrodynamic regime with $\lambda=2.5,\,h=0.01,\,q_\text{M}=10^{-6}$ referred to as \texttt{FID-I} in \citetalias{li22}. By reproducing this case under the new boundary condition framework, we verified that the modifications do not compromise the qualitative features reported previously. A detailed discussion and the results from this test run are provided in Appendix \ref{sec:app1} supporting the validity of the modified boundary conditions.}

\begin{table}[h]
    \centering
        \caption{Details of the size of the computational domain and parameters used for different simulation setups.}
    \begin{tabular}{|c|c|c|c|c|c|c|}
        \hline
        \multirow{2}{*}{{$\lambda$}} & \multicolumn{3}{|c|}{{Domain size}} & \multirow{2}{*}{{$\Omega_\text{k}$}} & \multirow{2}{*}{{$p$}} & \multirow{2}{*}{\textbf{$c_\text{s}$}} \\ \cline{2-4}
                                          & {$x$} & {$y$} & {$z$} &                         &                         &                         \\ \hline
        2.5  & (-12,12)  & (-12,12)  & (-5,5)  & 0.253  & 0.250  & 0.632  \\ \hline
        5.0 & (-20,20) & (-20,20) & (-10,10) & 0.089 & 0.125 & 0.447 \\
        \hline
        7.5  & (-20,20)  & (-20,20) & (-10,10) & 0.048 & 0.083 & 0.365 \\
        \hline
    \end{tabular} 
    \label{tab:param}
\end{table}

   \begin{figure*}[h]
   \centering
   \includegraphics[width=\textwidth]{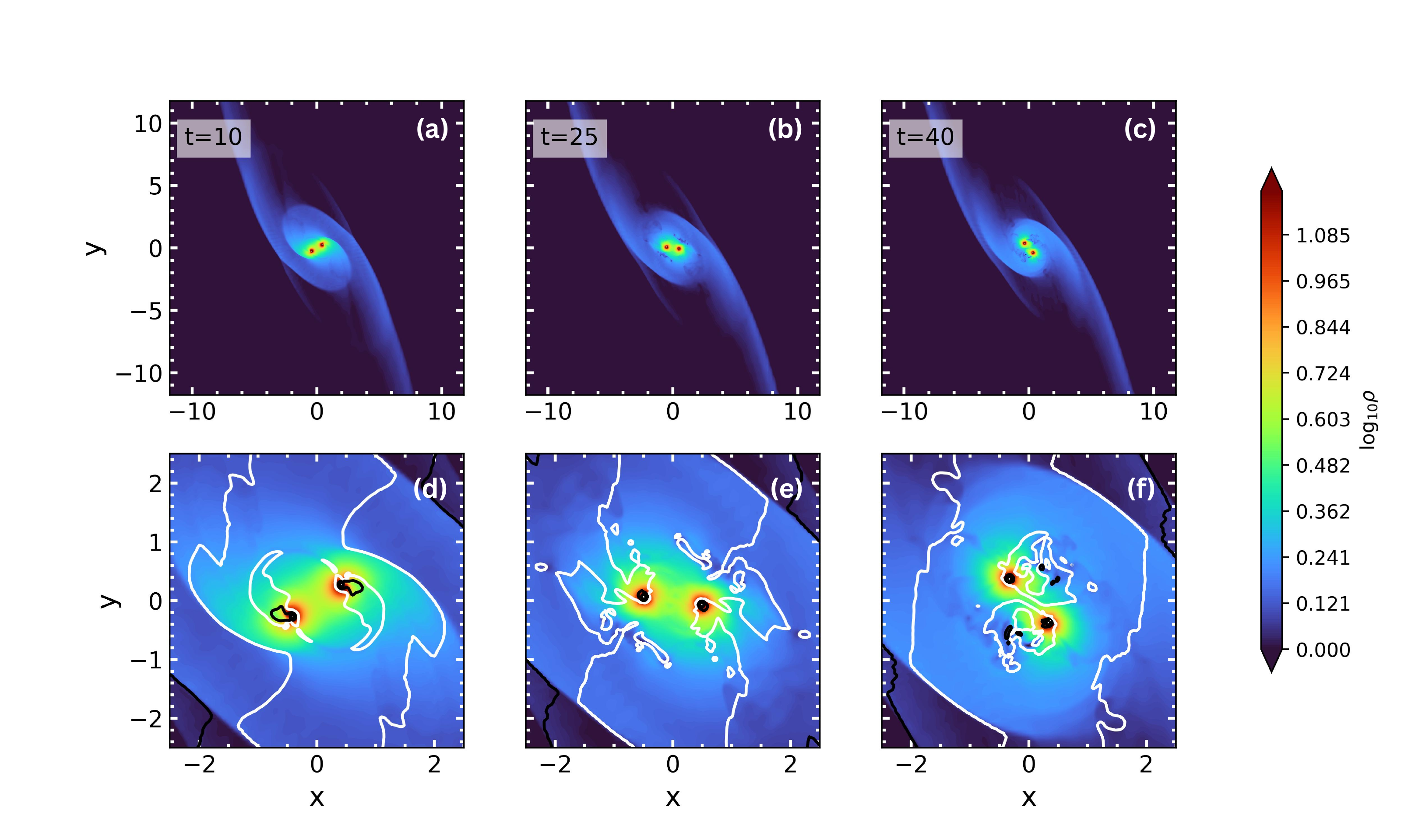}
      \caption{2D slices for $\log_{10}\rho$ at $z=0$ plane are plotted at different times to show the overall morphology of accretion flow. The bottom panels show the zoomed region near the BBH to highlight the disk structure. The line contours for sonic Mach number $\mathcal{M}$ are also plotted in the bottom panels, white lines show $\mathcal{M}=1.$}
         \label{fig:plane_xy}
   \end{figure*}
\section{Results} 
We present the outcomes of three different simulation runs, each characterized by a distinct value of the parameter 
 $\lambda$.  These simulations allow us to systematically investigate how varying $\lambda$ influences the local disk conditions of AGN disk and, subsequently dynamics of accretion and outflow structures around the binary. We begin by examining the case with $\lambda=2.5$, which prominently exhibits the formation of well-collimated outflows. In this scenario, we analyze the detailed morphological features associated with both the accretion flow and the resulting outflows. Subsequently, we consider simulations with progressively higher values of $\lambda$. As we shall see these runs reveal a noticeable trend: as $\lambda$ increases, the outflow activity becomes increasingly suppressed, indicating a weakening of the mechanisms responsible for driving and maintaining strong outflows at higher values. 
\label{sec:res}
\subsection{Dynamics in equatorial plane}
In Fig. \ref{fig:plane_xy} (a)-(c), we present 2D slices of  $\log_{10}\,\rho$ during the early stages of evolution. At this phase, the evolution is purely a hydrodynamical process governed by the interplay between the gravitational torques exerted by the binary and the resulting pressure gradients within the flow. 
  
The accreting BBHs draw in material through well-defined tidal accretion streams. This results in the formation of distinct mini-disks around each black hole, referred to as circum-single disks (CSDs).
Beyond the immediate vicinity of the black holes, the large-scale spiral arms extend outwards, reaching the azimuthal boundaries along the direction of the shear flow. These bow shocks arise due to the supersonic nature of the gas inflow (as in \citetalias{li22}). The formation of these spiral arms is mostly influenced by two parameters, namely $\Omega_{\text{K}}$ and $c_\text{s}$. {Higher $\Omega_{\text K}$ produces more tightly wound spiral arms, while larger $c_\text{s}$ results in broader, more diffuse structures. The detailed influence of these parameters is presented in the Appendix \ref{sec:app2}.}

   \begin{figure*}[h]
   \centering
   \includegraphics[width=\textwidth]{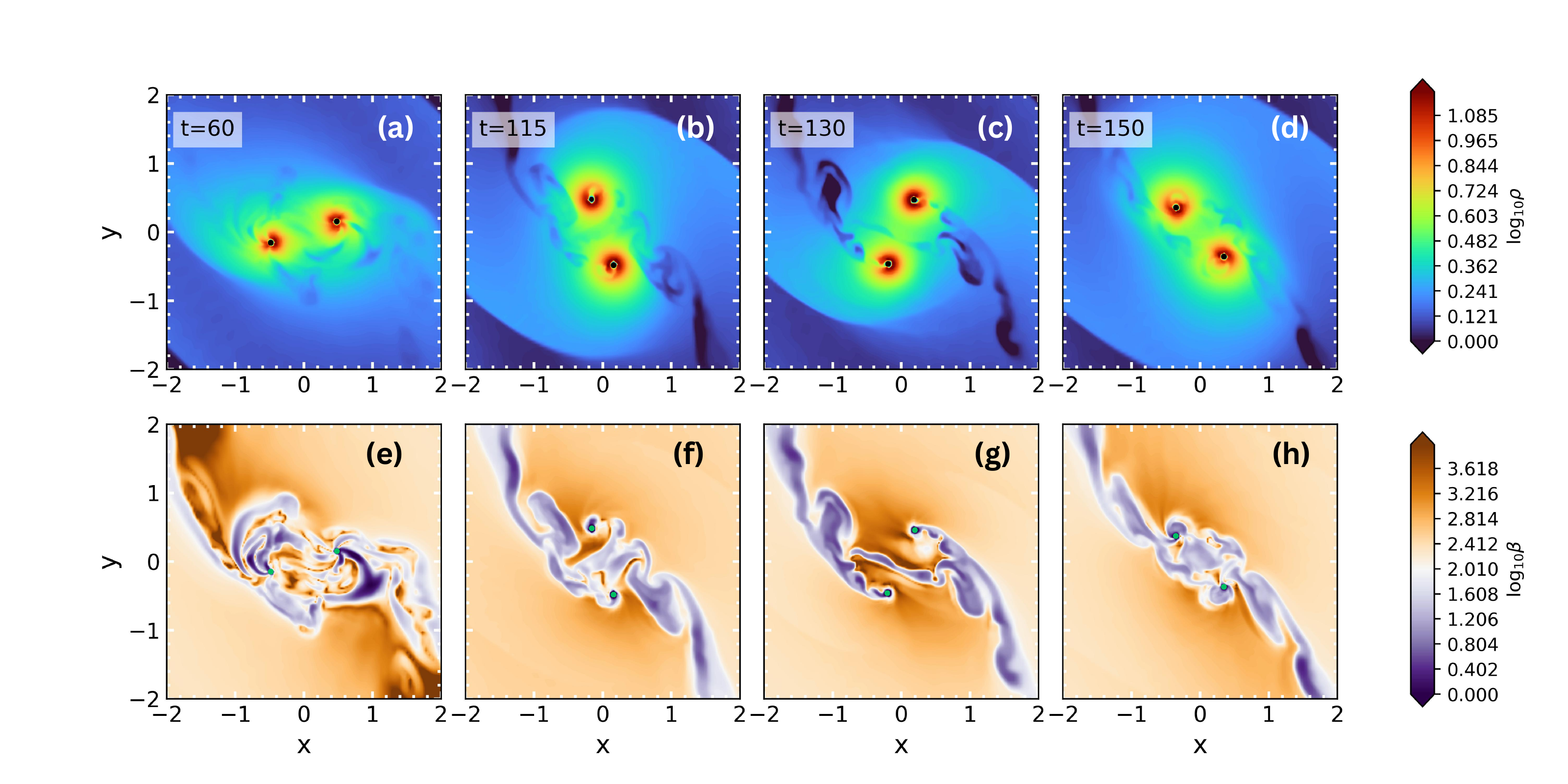}
      \caption{Density and plasma beta slices at $z=0$ plane in $\log_{10}$ scale are plotted at various epochs to show magnetic field accumulation in CSDs. The sink regions are masked by the black and green circles in the top and bottom rows for better illustration.}
         \label{fig:mag_plane_xy}
   \end{figure*}
It is important to emphasize that the size of the computational domain is sufficiently large to fully contain these spiral arms. This ensures that the flow structure remains physically consistent and is not affected by artificial truncation effects at the boundaries along the $x$-direction \citep{wh24}. 
Furthermore, the implementation of a standard outflow boundary condition along the $x$-direction prevents unphysical reflections or distortions at the domain edges. As a result, the global flow properties remain well preserved, allowing for an accurate representation of the large-scale accretion dynamics. The structure of the spiral arms remains largely unchanged during the initial phases of evolution as the overall mass distribution evolves gradually. The most intense dynamical changes are confined to the region inside the Hill sphere of the BBH ($R_\text{H}=2.5 a_\text{b}$, in this case), where strong gravitational interactions and differential accretion lead to rapid variability. Within this domain, the gravitational potential of the BBH dominates the gas dynamics, resulting in complex phenomena such as shock formation, angular momentum exchange, and sharp fluctuations in local density. 

In the bottom panels (d)-(f) of Fig. \ref{fig:plane_xy}, we also plot the contour lines of the sonic Mach number, defined as $\mathcal{M} = v/c_\text{s}$ where $v$ is the local flow velocity and $c_\text{s}$ is the sound speed of the gas, to highlight the transitions between different flow regimes. {We have plotted the contours for $\mathcal{M}=1$ with white lines, which separate supersonic and subsonic flows, highlighting the locations of shock transitions. The accreting gas undergoes a shock transition before being captured into CSDs around individual black holes.} This transition is critical for regulating the mass accretion process, as the shocks dissipate kinetic energy and allow the gas to settle into bound orbits around the black holes.

Overall, the early stages of evolution are non-magnetized accretion, where the hydrodynamical effects of shock formation, shear flow interactions, and acceleration induced by the gravitational pull of the binary govern the morphology of the accretion structures. 
At the later stages, magnetic fields embedded in the accreting gas accumulate around the binary black holes, leading to additional modifications in the dynamics of the system.

Once the magnetic field is injected, it starts to accumulate in regions near BBH, and inhomogeneous structures develop around CSDs, as illustrated in Fig.~\ref{fig:mag_plane_xy}(a)-(d). The gravity of binary and shock fronts compress the plasma, which in turn leads to the enhancement of the magnetic field. To further investigate the interplay between gas and magnetic fields, we analyze the evolution of the plasma $\beta$, which quantifies the relative dominance of gas pressure compared to magnetic pressure. The bottom panels (e)-(h) of Fig.~\ref{fig:mag_plane_xy} track changes in $\beta$ over time, providing insight into how the accretion flow transitions from a gas-pressure-dominated regime to one increasingly controlled by magnetic forces.

At $t=60$, in Fig. \ref{fig:mag_plane_xy}(e), one can see that the individual CSDs have accumulated significant magnetic fields, creating the regions of low $\beta$ . These low $\beta$ regions are typically associated with reduced density. The presence of strong magnetic fields also disrupts the uniformity of the plasma, giving rise to turbulent and filamentary structures. The accumulation of enough magnetic pressure leads to eruption-like events. {The buildup of magnetic pressure gives rise to eruption-like events, which are identified once the accumulated pressure becomes sufficient to trigger the release of magnetic flux bundles into the disk. As accretion onto the black holes continues, these flux tubes are expelled. Their ejection is characterized by localized regions of lower plasma $\beta$, and a clear decoupling from the black holes can be noticed in Fig. \ref{fig:mag_plane_xy}(f)-(g), plotted at times $t=115$ and $t=130$.}
The continued existence of accretion onto the binary through spiral arms keeps feeding the matter and magnetic field to CSD despite ongoing turbulence and small-scale structural evolutions, resulting in the accumulation of the magnetic field, as shown in Fig. \ref{fig:mag_plane_xy}(h), after the eruption cycle. \\

\begin{figure}
\centering
\includegraphics[width=0.9\columnwidth]{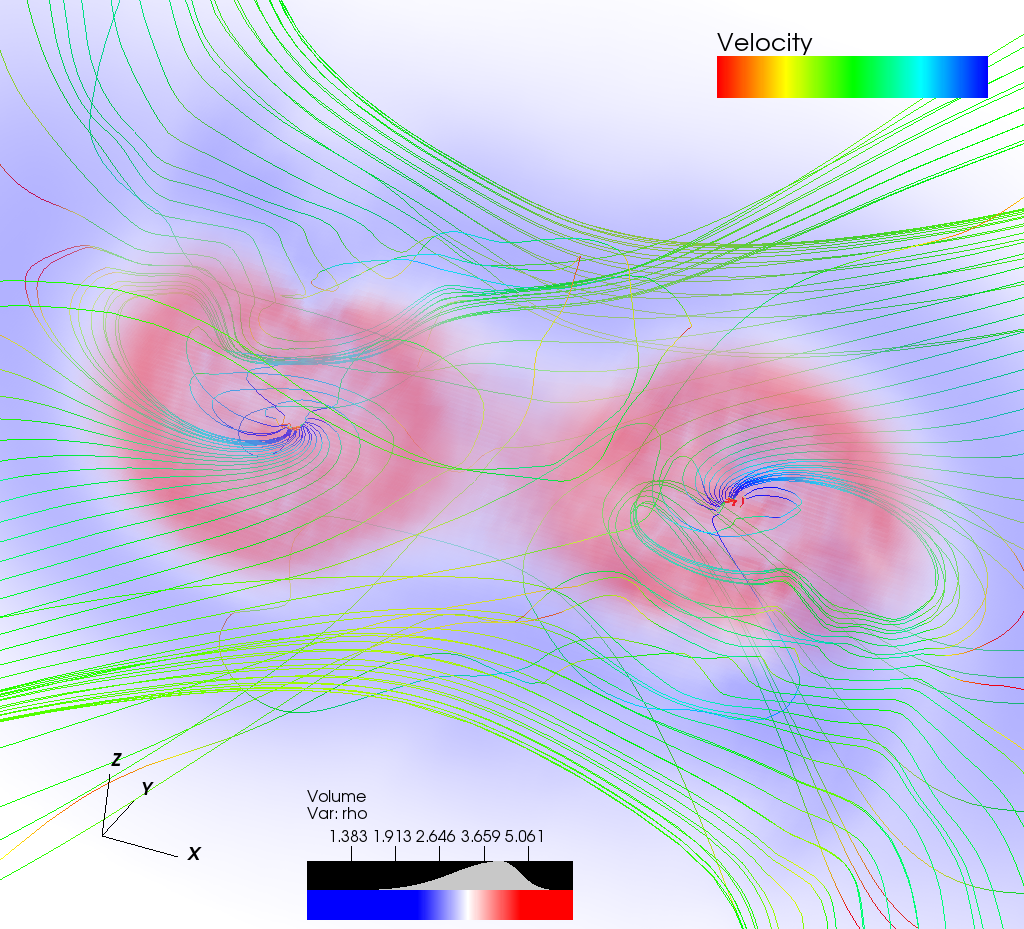}
\caption{Three-dimensional volume rendering of density overlaid with velocity streamlines, illustrating the spatial distribution of density structures of CSDs and the corresponding flow dynamics.}
\label{fig:3d_morph}
\end{figure}

   \begin{figure*}
   \centering
   \includegraphics[width=\textwidth]{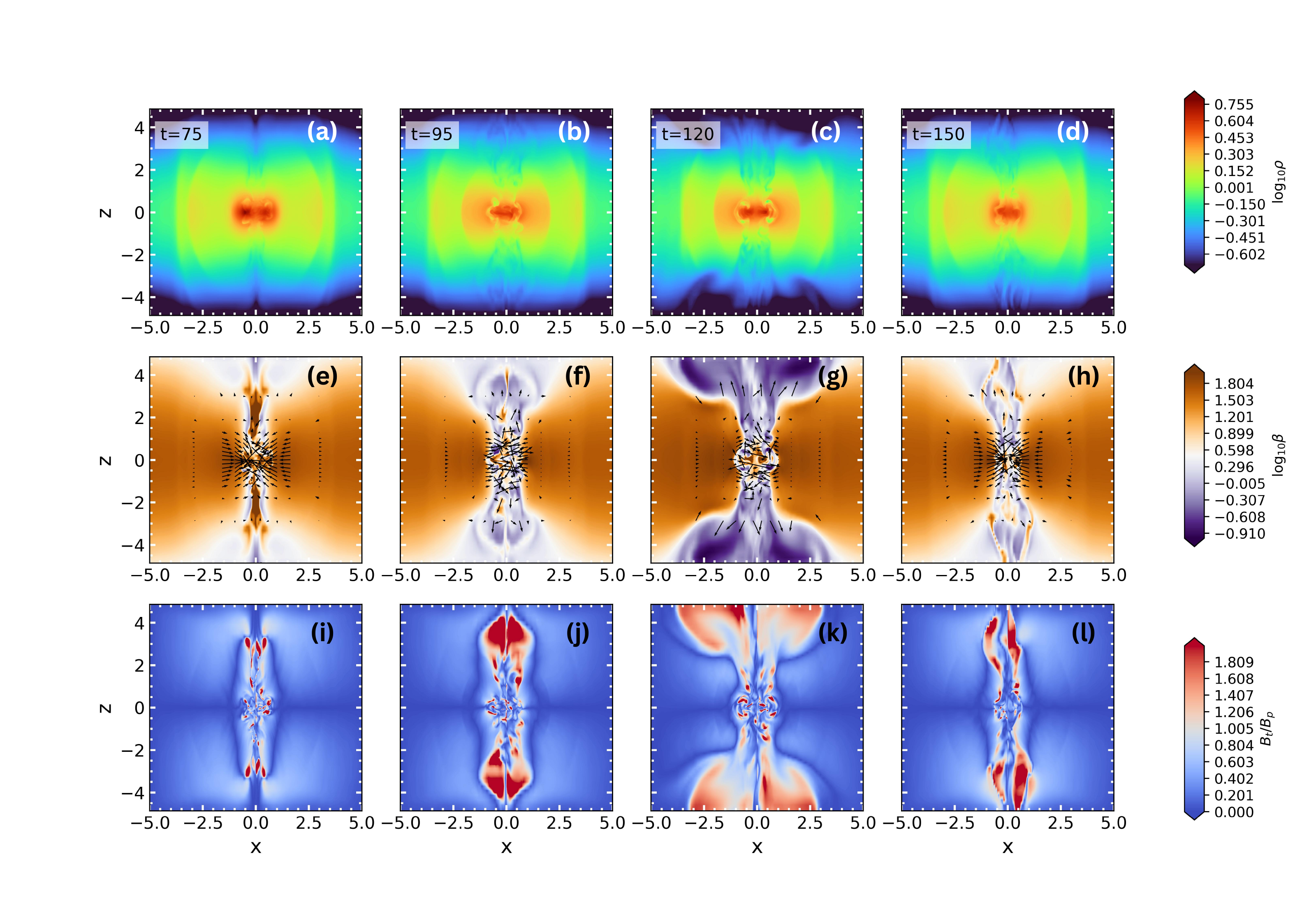}
      \caption{2D distribution of $\log_{10}\rho$ and $\log_{10}\beta$ along with velocity vectors are plotted in the $x-z$ plane ($y=0$) at different time stamps mentioned in the panels. The bottom row shows the ratio of toroidal and poloidal components of the magnetic field.}
         \label{fig:vertical_slice2}
   \end{figure*}

\begin{figure*}
   \centering
   \includegraphics[width=0.8\textwidth]{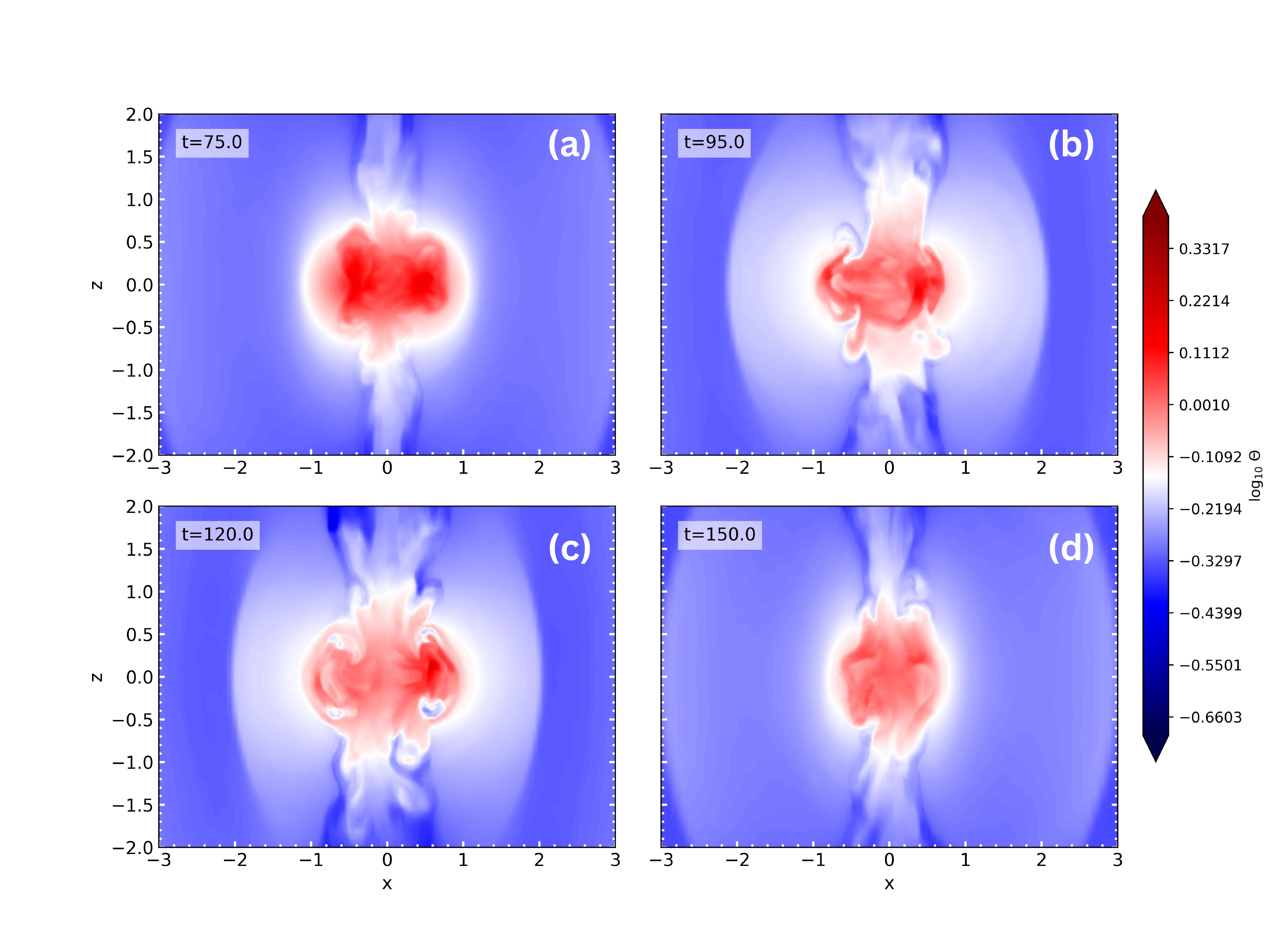}
    \caption{2D snapshots for $\log_{10}\Theta$, ($\Theta=p/\rho$) in $x-z$ plane are plotted at different time stamps to highlight the vertical structures of the thick CSDs.}  
    \label{fig:theta_temp}
\end{figure*}
    
In Fig. \ref{fig:3d_morph}, we present a three-dimensional volume rendering that illustrates the morphology of the individual CSDs, highlighting their density structure. Superimposed velocity streamlines depict the large-scale flow topology and are color-coded by velocity magnitude, with red indicating lower values and blue representing higher ones. The streamlines form a characteristic U-shaped (horseshoe) pattern as they wrap around the black hole before plunging inward. Notably, the flow exhibits a columnar structure along the horseshoe turn, where streamlines undergo a sharp drop in altitude during their transition. These features are consistent with previous numerical investigations \citep{Fung_2015}.  

\subsection{Outflows}
\label{sec:outflow}
The top panels~(a)-(d)~of Fig.~\ref{fig:vertical_slice2} illustrate the behavior of accreting material in the $x-z$ plane at $y=0$ by showing logarithmic density slices at different time steps. These snapshots provide insight into the vertical structure of the individual CSDs surrounding each black hole. A noticeable feature of these disks is their puffed-up structure, which closely resembles a thick torus. This characteristic shape emerges as a direct consequence of the way infalling gas interacts with the strong gravitational potential of the black holes.
As gas with angular momentum from the surrounding environment is drawn toward the binary, it undergoes a shock transition; as a resultant process, the kinetic energy is rapidly converted into thermal energy \citep{mr96,cc11,scl21,dcj24}. This sudden energy transformation leads to intense heating of the plasma, significantly raising its internal pressure.\\

In Fig. \ref{fig:theta_temp}, we plot the 2D slices of $\Theta=p/\rho$, which is a measure of temperature ($p/\rho\propto T$). 
The heated material expands vertically, giving rise to the thick, puffed-up structures observed in our simulations. Despite their expanded nature, these hot tori do not completely block the inward flow of gas. Instead, accreting material is funneled through narrow openings, creating well-defined channels that guide the inflowing gas toward the black holes. The density and temperature evolution of these torus shaped structures, shown in Fig. \ref{fig:vertical_slice2} and \ref{fig:theta_temp} indicates that these are not steady-state tori. The precise morphology of these structures is shaped by the dynamical balance between thermal pressure, magnetic forces, and the angular momentum of the infalling material \citep{g20,dvf22,dm25}. The combined effects of these forces determine the stability and shape of the accretion flow around the binary. \\

During the earlier epochs of simulation, the system exhibits relatively high $\beta$ values, indicating that gas pressure is the dominant force shaping the dynamics. However, as time progresses, a significant decrease in $\beta$ is observed, meaning that magnetic pressure becomes increasingly influential. This shift toward a magnetically dominated state is driven by two key mechanisms that lead to magnetic field accumulation. The first mechanism is the advection of magnetic flux by the accretion flow. As matter spirals inward toward the black holes, it carries along the frozen-in magnetic field, effectively dragging it inward. This inward transport of magnetic flux naturally leads to an enhancement of the field strength near the binary. As more magnetic flux accumulates, the magnetic pressure builds up, contributing to a dynamically evolving environment. The second mechanism is field line stretching due to the binary’s orbital motion. The orbital motion of the black holes around their common center of mass plays a crucial role in amplifying the magnetic field. \\

One of the most important consequences of magnetic field accumulation is the emergence of a magnetically dominated funnel region, clearly visible in the middle panels (f) and (g) of Fig.~\ref{fig:vertical_slice2}. This funnel forms due to the accumulation of strong magnetic pressure gradients, which exert outward forces on the surrounding material. As a result, matter is expelled along the rotation axis, carving out a low-density channel. This magnetically dominated funnel is a defining feature of accreting black hole systems and is {correlated} with the launching of strong outflows. The high degree of magnetization in this region indicates that magnetic forces exert a dominant influence over the dynamics of the outflows, effectively shaping their structure and behavior.\\

In addition to enhancing magnetic pressure, the accretion process also leads to the generation of a toroidal component of the magnetic field. This transformation occurs due to the rotational motion of the gas in the CSD, which twists and stretches the initially vertical field lines into azimuthally oriented structures. The bottom panels of Fig.~\ref{fig:vertical_slice2} illustrate the evolution of the ratio between the toroidal and poloidal components of the magnetic field over time. Initially, when the simulation begins, the magnetic field is purely vertical, meaning there is no toroidal component. However, as accretion proceeds, the rotation of accreting gas twists the field lines, progressively generating a toroidal component. \\

At $t = 75$, the ratio $B_\text{t} / B_\text{p}$ is still quite low, indicating that the toroidal component remains weak. The plasma $\beta$ is relatively high near the binary, meaning that gas pressure is still dominant. The velocity vectors in Fig.~\ref{fig:vertical_slice2}(e) are all directed inward, signifying accretion onto the black holes. As this toroidal component grows in strength, magnetic buoyancy effects become increasingly significant. Magnetic buoyancy can drive the formation of rising poloidal loops, where magnetic field lines arch upward due to pressure imbalances. This phenomenon has been studied extensively in the context of accretion disks around black holes, binary stars, and planets \citep{bp82, tom10, g20, m24}. By $t = 95$, a significant buildup of magnetic pressure is observed, leading to a lower value of $\beta$. The increase in magnetic pressure counteracts gravity and opposes the accretion, with some vectors beginning to point outward. This suggests the early stages of outflow formation as the magnetic field becomes more dynamically significant. The toroidal component of the field is now well-developed, reinforcing magnetic pressure gradients. Around $t = 120$, the system transitions into a state with a fully developed funnel region and strong outflows. However, magnetic collimation is not the only collimation mechanism involved; the vertical structure of the disk itself also plays a crucial role in confining the outflow. The thickened toroidal shape of the inner disk effectively acts as a natural barrier, restricting lateral expansion and aiding in the formation of a well-collimated jet. Accretion onto the black holes significantly weakens as the magnetic pressure has grown strong enough to suppress further accretion. The morphology of these outflows closely resembles that of the well-known ``magnetic tower jet'' \citep{llgf06}. \citet{nll06} demonstrated the formation of such jets in a stratified atmosphere under a Newtonian gravitational potential, where matter is injected along with a magnetic field dominated by toroidal flux over poloidal flux. This configuration leads to the development of a low-density cavity characterized by low plasma $\beta$ and high Alfvén speed. In our simulations, the magnetic field is advected inward with the accreting gas, while the toroidal component is generated through the combined effects of accretion and rotation. The resulting contour plots of $\beta$ and $B_t/B_p$ show that the required conditions are satisfied in these regions, consistent with the magnetic tower jet scenario.

As the system continues to evolve, the strength of the jet activity is not constant but exhibits temporal variations. This variability arises because the magnetic energy stored in the system is gradually depleted over time. As a result, the jet weakens, and $\beta$ begins to increase again, signaling a return to a more gas-pressure-dominated state. This transition is clearly visible in Fig.~\ref{fig:vertical_slice2}(h), where an increase in $\beta$ correlates with a decrease in the vertical velocity component, $v_\text{z}$, indicating a slowing down of the outflow. This behavior suggests that jet activity in such systems is not continuous but rather transient, with phases of strong magnetic launching followed by periods of weaker jet production. These results suggest that such systems may exhibit intermittent jet activity, potentially influencing their observational characteristics over long timescales. {This phenomenon is similar to that of single accreting black holes \citep{nia03,tnm11,porth21,dvf22,salas24,moscri25}, where the accumulation of net magnetic flux partially inhibits accretion, leading to quasi-periodic cycles of flux accumulation and ejection. However, whether the periodic behavior can persist in the BBH case depends critically on the lifetime of the binary before merger. If the binary merges before accumulating sufficient net magnetic flux to sustain multiple jet-launching cycles, the observed jet activity could be a one-time event rather than a recurring phenomenon. Thus, the duration of the binary’s inspiral sets an upper limit on the possible number of jet activity episodes, ultimately influencing the long-term observational signatures of these systems \citep{ressler25,mw25,mpb25}.}

\subsection{Field line topology}

\begin{figure*}
   \centering
   \includegraphics[width=0.6\textwidth]{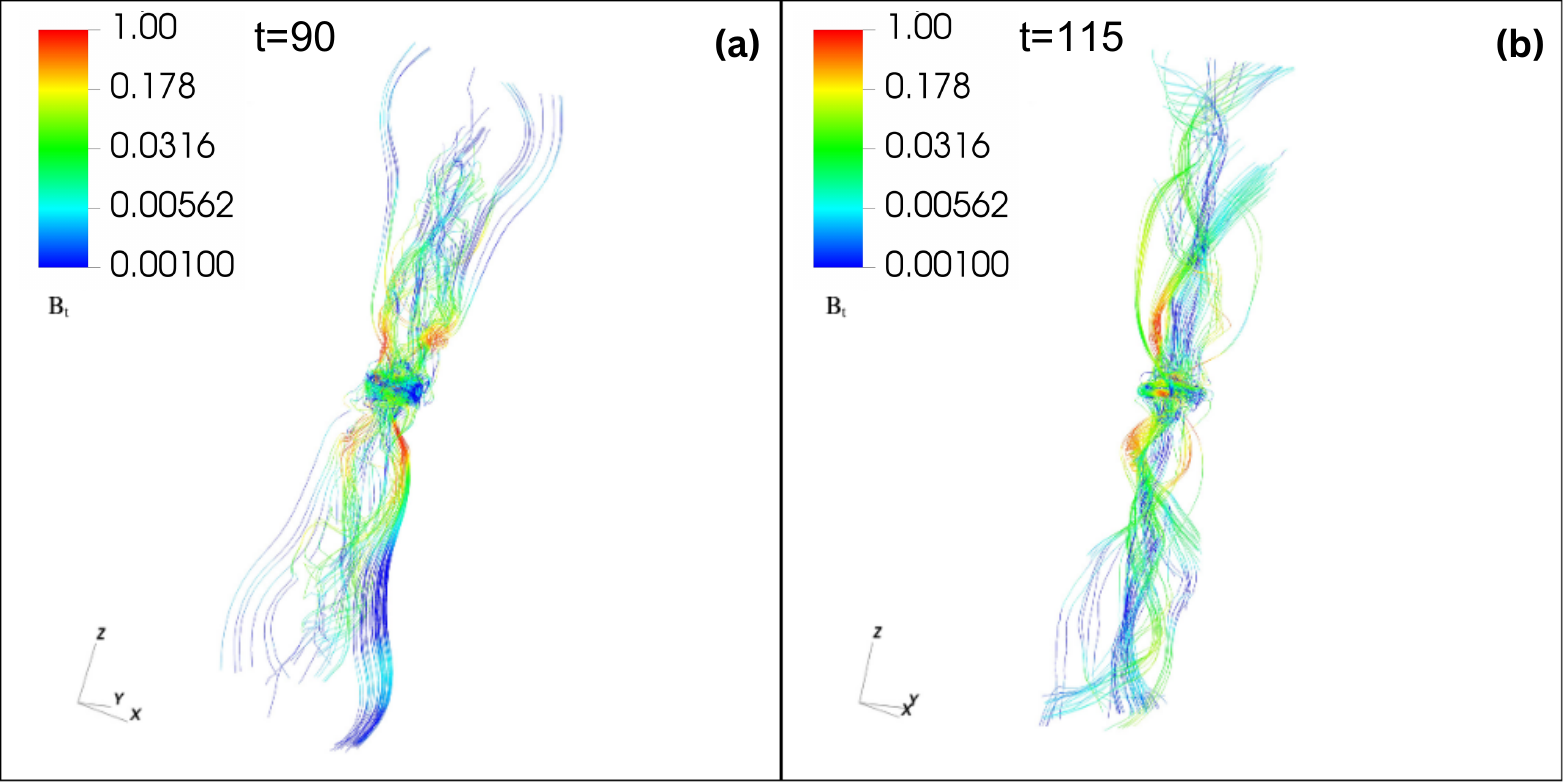}
      \caption{Three-dimensional view of magnetic field lines. The lines are color-coded with the magnitude of $B_\text{t}$. $B_\text{t}$ and $t$ are in code units.}
         \label{fig:B_field}
   \end{figure*}

To better illustrate the evolution of the magnetic field, we present a visualization of magnetic field lines in Fig. \ref{fig:B_field}. This figure depicts how rotation influences the structure of the magnetic field. The magnetic field lines are color-coded with the strength of the toroidal component $B_t$.
The field lines show a significant deformation from the initial pure vertical topology. At $t=90$ we can see the enhancement of $B_t$ near the central region, due to the formation of CSDs and orbital motion of BBH around the center of mass; the field lines show a twisted structure. Initially, this twisting effect is confined near the central region. The generation of the toroidal component ($B_\text{t}$) induces a pinching force, causing the field lines to readjust into an hourglass-like configuration \citep{kud02, mach11}. As discussed in Section \ref{sec:outflow}, the presence of $B_{t}$ leads to magnetic buoyancy, facilitating the vertical transport of stored magnetic energy. This process results in an enhancement of $B_t$ strength near the edges of the z-boundary, as seen in the right panel at $t=115$.

\subsection{Dependence on distance from SMBH}
\label{sec:no_jet}

   \begin{figure*}
   \centering
   \includegraphics[width=\textwidth]{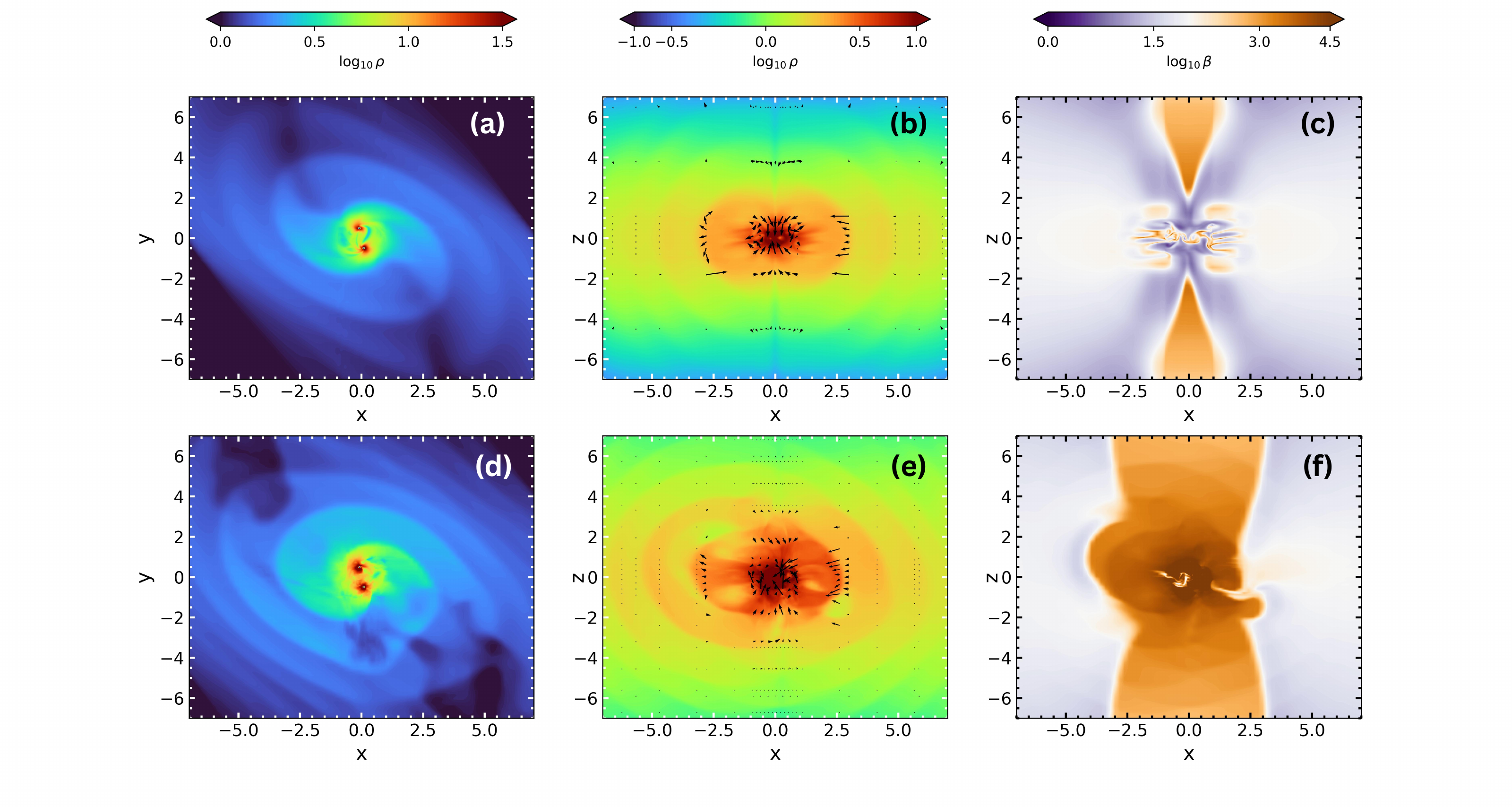}
      \caption{The 2D snapshots for $\log_{10}(\rho)$ in $z=0$, $y=0$ plane and $\log_{10}\beta$ are plotted at $t=96$ for $\lambda=5$ (top row) and $\lambda=7.5$ (bottom row)}.
         \label{fig:vertical_slice_lambda_5}
   \end{figure*}

A closer look at Eq. (\ref{eq:param3}) indicates that for the fixed values of $a_\text{b}$ and $q_\text{M}$, higher values of $\lambda$ mean that BBH is seeded at higher radial distances from SMBH. For higher values of $R$, the background gas has a lower velocity (see Eq. \ref{eq:back_wind}). The change in $\lambda$ also modifies the sound speed $c_\text{s}$, meaning that the local thermodynamic conditions are different for the BBH system. To investigate this effect in more detail, we perform two additional simulations in which we take $\lambda=5, 7.5$, corresponding to a regime where the infalling gas has relatively lower azimuthal velocities. By varying only this parameter while keeping the same initial $\beta$, we ensure that both simulations are directly comparable in terms of the magnetic energy available at the onset of accretion. Additionally, the change in $\lambda$ while keeping $h$ and $a_\text{b}$ constant (see Eq. \ref{eq:param3}) means we are studying the same BBH system seeded on AGN disk at higher distances.\\ 

In Fig. \ref{fig:vertical_slice_lambda_5}, we show the contour plots for these simulations. The top row is for $\lambda=5$ and the bottom row for $\lambda=7.5$. We plot the 2D distribution of $\log_{10}\rho$ in the $z=0$ plane and the $y=0$ plane in the first and second columns of figure, respectively, and $\log_{10}\beta$ for the $z=0$ plane is plotted in the third column of the figure. By comparing these results with those shown in Fig. \ref{fig:plane_xy}, we observe a clear trend: as the value of $\lambda$ increases, the overall flow pattern around the BBH becomes increasingly turbulent. In particular, the spiral shock fronts become more pronounced. This transition is driven by the increase in the ratio of the binary orbital velocity to the background sound speed, $v_\text{b}/c_\text{s}$, as described in Eq. (\ref{eq:param2}). Since this ratio increases with $\lambda$, the Mach number associated with the orbital motion of the binary also becomes larger. A higher Mach number implies that the flow transitions from being subsonic and relatively smooth to supersonic and turbulent. This behavior is consistent with previous findings in the literature, such as those discussed in \citet{lilai23}, which highlight the correlation between higher Mach numbers and the onset of turbulence in accretion flows. Additionally, we also notice that the circumbinary flows and the CSDs become denser with a higher value of $\lambda$. \\

The velocity vector fields overlaid on the density distribution show a predominant inward motion of the gas towards the BBH, with no visible signs of outflows or jets being launched from the system. This lack of outflow signatures can be understood in the context of the BBH’s environment. When the binary is embedded in the outer regions of AGN disk, the surrounding gas often possesses very low angular momentum. In such a scenario, the gas cannot form a rotationally supported disk and instead undergoes nearly radial infall onto the binary system \citep{kaaz23}. \\
{A larger value of $\lambda$ indicates stronger binding between the binary and the surrounding gas (i.e., a larger Hill radius relative to the binary separation, giving the binary a greater gravitational reach over nearby disk material), which prevents the formation of a low-density funnel region (see Fig \ref{fig:vertical_slice_lambda_5}-(b) and (e)) and suppresses the outflow.}

The accumulation of matter enhances the magnetic field in these runs as well, but there is no ordered distribution of low $\beta$ regions, unlike $\lambda=2.5$ case where the low $\beta$ channel is carved in a funnel-like bipolar geometry. We also checked for the distribution of $B_t$, which also remains weak in these runs, resulting in no outflow activity. \\

\subsection{Physical Scaling of Simulation Parameters}
\label{sec:scaling}
{
To connect our dimensionless simulation results with astrophysical systems, we need to specify the binary separation and mass to calculate the parameters in physical units, as we have normalized all the units with respect to the binary parameters. Assuming a binary with total mass $ m_\text{b}=100M_{\odot}$ and separation of $100$ AU, our $\lambda=2.5$ simulation means we are studying a binary embedded inside the disk of an AGN of BH mass $10^8M_{\odot}$ at a distance of $R=3.77\times10^{17}\rm cm$. The orbital period of the binary system is $100.05$ years. We have adopted a standard steady $\alpha$ \citep{ss73} disk model for AGN. For the density   
if we adopt a fiducial accretion rate of $0.1M_\text{Ed}$ (where $M_\text{Ed}=1.44 \times 10^{17}M_\text{BH}/M_\odot\,\rm g/s=1.44 \times 10^{25}\,\rm g/s$ is the Eddington accretion rate), typical mass density at the equatorial plane turns out to be the order of $\rho_0\simeq10^{-12}\rm g/cm^3 $, and the magnetic field strength is $3.93\times 10^{-1}\rm G$.}\\

We also calculate net mass flux density, defined as 
\begin{equation}
\Phi=\sum\rho(\mathbf{v}\cdot\hat{n})dA/\mathbf{S}    
\label{eq:mflux}
\end{equation}

The summation is performed over all the faces of a cubic volume. The side length of the cubic volume is chosen to be equal to $\lambda$, which fully encloses the region bounded by the Hill sphere (see Eq. (\ref{eq:param2})), ensuring that it captures the relevant dynamics of mass exchange occurring around the binary system, $\mathbf{S}$ is the total surface area of the cube. The value of $\Phi$ serves as a diagnostic tool to evaluate the relative dominance of accretion versus outflow processes within the simulated environment. A positive value of $\Phi$ indicates a net outflow of matter from the volume, while a negative value signifies net accretion onto the central object. \\

\begin{figure}
   \centering
   \includegraphics[width=\columnwidth]{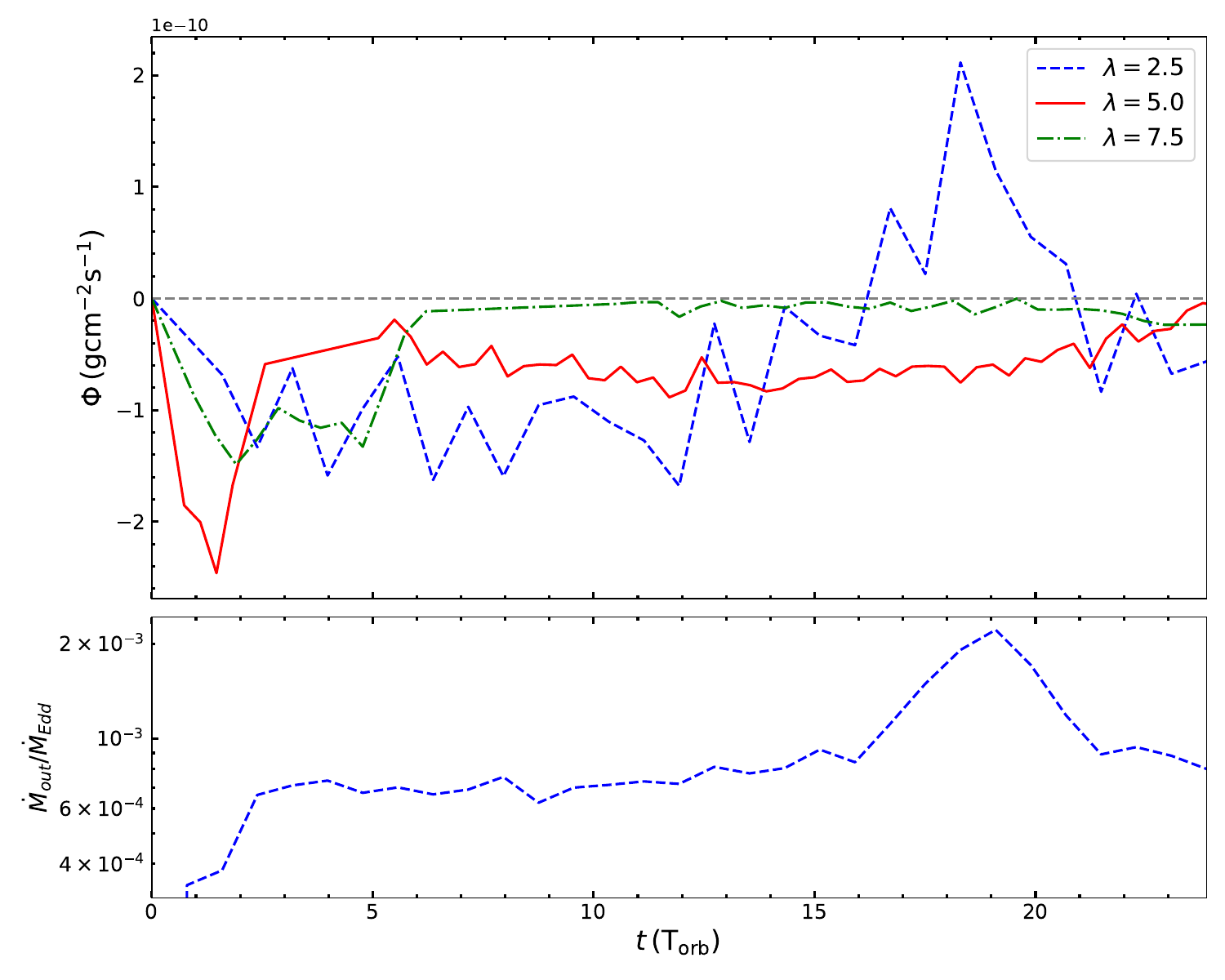}
    \caption{Time series for mass flux density computed using Eq. (\ref{eq:mflux}) for $\lambda=2.5,5,7.5$ runs. Positive values of $\Phi$ indicate a net outflow of matter through the volume. The bottom panel shows the mass outflow rate in Eddington units through the upper and lower faces of the computational domain. The time is measured in units of the binary's orbital period.}  
    \label{fig:mass_flux}
\end{figure}

The temporal evolution of $\Phi$ is illustrated in Fig.~\ref{fig:mass_flux}, where the variation of the mass flux density is shown as a function of time for all simulation runs under consideration. The dotted blue line in the figure corresponds to the case with $\lambda = 2.5$. In this run, as material from the surrounding envelope begins to accrete onto the binary, $\Phi$ becomes negative, signifying that more mass is entering the control volume than leaving it. Although it does not decrease monotonically, it maintains an overall negative trend during the early stages of the simulation. This behavior confirms that accretion dominates the early evolution of the system for this particular value of $\lambda$.\\

Around time $t \sim 16 T_{\rm orb}$, a noticeable transition in $\Phi$ is observed: the value of mass flux density crosses zero and becomes positive. This change indicates the onset of a significant outflow component. This transition correlates well with the emergence of a jet-like structure seen in the vertical slice of density and velocity shown in Fig.~\ref{fig:vertical_slice2}. The value of $\Phi$ continues to increase and reaches a peak around $t \sim 19T_{\rm orb}$, suggesting a phase of enhanced outflow activity. This peak is attributed to the release of stored magnetic energy, which temporarily powers the outflow. However, as the magnetic reservoir depletes, the outflow rate starts to drop down. Consequently, the value of $\Phi$ declines, and the simulation transitions into a phase where accretion once again dominates over outflow. \\

In contrast, the simulation runs with $\lambda = 5$ and $\lambda = 7.5$, represented by the solid red and dash-dotted green lines, respectively, exhibit a qualitatively different behavior. For these higher values of $\lambda$, $\Phi$ remains consistently negative throughout the entire duration of the simulations. This indicates that, in these cases, the system experiences continuous net accretion, with no phases where outflows dominate. Nevertheless, both curves display several localized peaks and fluctuations, implying that the accretion rate is not steady but rather variable in time. 

{In the bottom panel of Fig. \ref{fig:mass_flux}, we have plotted the mass outflow rate (in Eddington units) through  $\pm z$ boundaries of the computational domain for the case $\lambda=2.5$, the outflow rate has a maximum value of $\sim 1.5\times10^{-3}M_\text{Ed}$.}

We also compared these profiles with the higher resolution run and found that they agree well. For the $\lambda=2.5$ case, the Pearson's correlation coefficient turned out to be $ r=0.9739$ and the percentage agreement based on $r^2$ was $94.85\%$.

\section{Conclusions}
\label{sec:con}
\subsection{Summary}
We have studied the evolution of BBHs in a magnetized AGN disk in an ideal MHD framework using a local shearing box approach. We have studied binaries in a fixed circular orbit with equal masses, embedded in the equatorial plane of a standard Keplerian disk of an AGN. In our simulations, we take a fixed mass ratio between BBH and SMBH $q=m_\text{b}/M=10^{-6}$ and a fixed disk aspect ratio $h=0.01$ and vary the ratio between the binary Hill radius and binary separation $\lambda$ for different simulations to investigate the morphological differences in accretion properties of the BBH system. We use modified boundary conditions in the shearing box framework, which allow us to inject matter continuously such that the computational box has a reservoir of sufficient mass flux even if the matter is being lost through vertical outflows. We use non-uniform grids with a higher resolution near the center of the box to capture the features associated with the BBH system more accurately. Our findings demonstrate that the evolution of binaries embedded in accretion disks is responsible for outflow generation, provided the magnetic fields are dynamically important. These outflows are well collimated and strong enough to reach the vertical edges of the AGN disks and leave the computational domain. The generation of outflows is not ubiquitous but depends on the localization of the binary on the AGN disk, even if the binary parameters are kept constant. This is because the different regions of the disk provide different background conditions for BBH to evolve, and the transition to the state where the magnetic fields become dynamically important and generate outflows depends on the effect of the local environment. The BBHs embedded in the AGN disk are expected to have sufficient matter around them and potentially generate some EM radiation.  Yet, this radiation may be entirely blocked by the optically thick Keplerian disk that engulfs the binary. However, the formation of outflow is an interesting phenomenon that can mechanically clear the optically thick disk environment and help radiation to escape. The formation of outflow is closely associated with the velocity of the background gas, which influences the growth of the toroidal component of the magnetic field. If the background shear provided by the AGN disk is weak, the outflow rate drops down, and the flow morphology around the BBH is mostly accretion-dominated. Apart from the magnitude of background velocity, any change in the distance of the BBH from the SMBH also changes the local sound speed. This effect is clearly reflected in the structures formed close to BBH. The BBH system with a higher value of $\lambda$ shows more turbulent and chaotic accretion, leading to the formation of more prominent spiral shocks. Although we have not addressed this problem on the horizon scale of individual components of the BBH system, our simulations also show a similar morphology of well-collimated outflows similar to what has been reported by numerical relativity simulations \citep{pgll10,rts23}, despite the outflow generation mechanism in our simulations being different from the Blandford-Znajek mechanism \citep{bz77}. In addition to the formation of outflows, our results also show episodic accretion. These episodic accretion, eruption events, and subsequent ejection of plasmoids were also reported by \cite{most24}.                      

\subsection{Limitations and future work}
In our simulations, we have covered only the effect of one parameter $\lambda$ in the evolution. The other parameters, like $h/q_\text{M}^{1/3}$ and the ratio of Bondi radius to binary orbit, are two additional parameters that can have a strong influence on the evolution of the BBH system. It will be important to consider the different values of these parameters.
{
In our study, we assume that for fixed binary and SMBH masses, \( \lambda \propto R \), and thus varying \( \lambda \) corresponds to embedding the binary at different radial positions in the AGN disk. However, we emphasize that this simplified interpretation does not capture the full complexity of radial variation in realistic AGN disk environments. Physical quantities such as gas density, temperature, disk aspect ratio, and viscosity all vary significantly with radius. These variations are described in both classical \(\alpha\)-disk models \citep{ss73} and more complex, disk models \citep{sg03}. Additionally, the magnetic field strength is not constant but varies with radius due to its dependence on gas density and vertical structure. This variation in the magnetic field may be a key factor in the regulation of angular momentum transport and outflow launching efficiency.
Although our approach neglects other radial dependencies, it provides a controlled way to investigate the effect of local shear and the background flow on accretion and outflows around a binary system in a simplified but physically informative framework.   
Future work will extend this analysis by coupling binary dynamics to self-consistent, disk models, allowing us to examine how gas density, disk thermodynamics, and turbulence influence outflow properties in tandem with gravitational shear.
}
Apart from the limited parameter space explored in this study, the major caveat is that we have kept binaries in fixed orbit, so our results mainly address the dynamics of the pre-merger phase of the binary system. The general relativistic simulations of merging binaries \citep{rts23,kprs18} show that the jets persist through inspiral, merger, and post-merger. While our simulations are only restricted to fixed orbit configurations, they show the possibility of the formation of an outflow. In order to study the fate of the jet through the inspiral phase, we need a consistent analysis of the BBH orbit and the evolution of matter in the resulting gravitational field. This matter, while noteworthy, extends beyond the scope of the simulation framework adopted in this study. As such, it remains a topic for future research and will be explored in subsequent work, the results of which will be reported elsewhere.

\begin{acknowledgements}
BV acknowledges the support of the Max Planck Partner Group, established at IIT Indore. The computations in this work were performed using the
facilities at IIT Indore, Max Planck Institute for Astronomy cluster VERA, the computer cluster at the Nicolaus Copernicus Astronomical Center of the Polish Academy of Sciences (CAMK PAN), and Frontera at TACC through allocation AST20025. Frontera is made possible by NSF award OAC-1818253. This work was supported in part by National Science Foundation (NSF) Grants No. PHY-2308242, No. OAC-2310548 to the University of Illinois at Urbana-Champaign. MR acknowledges support by the Generalitat Valenciana Grant CIDEGENT/2021/046, by the Spanish Agencia Estatal de Investigaci\'on (Grant PID2021-125485NB-C21) and by the European Horizon Europe staff exchange (SE) program HORIZON-MSCA2021-SE-01 Grant No. NewFunFiCO-101086251. AT acknowledges support from the National Center for Supercomputing Applications (NCSA) at the University of Illinois at Urbana-Champaign through the NCSA Fellows program. BB and MB acknowledge financial support from the Italian Ministry of University and Research (MUR) for the PRIN grant METE under contract no. 2020KB33TP.
M\v{C} acknowledges the Czech Science Foundation (GAČR) grant No. 21-06825X.
\end{acknowledgements}

\bibliographystyle{aa} 
\bibliography{biblio}

\begin{appendix} 
\section{Comparisons with previously used numerical setups}
\label{sec:app1}
In the absence of magnetic fields, our simulations reduce to the hydrodynamical (HD) limit and are directly comparable to those of \citetalias{li22}, who also employed the shearing-box approximation to study BBHs embedded in AGN disks. The physical setup is largely similar: we use similar parameters $q_M = 10^{-6}$, $h = 0.01$ and $\lambda = 2.5$. Our study uses the \texttt{PLUTO} code (\citetalias{li22} use \texttt{ATHENA++}) and implements a $\gamma$-law equation of state with $\gamma = 1.6$, we make a notable departure in two respects: (1) the gravitational softening length, which in our case is set slightly larger to prevent extreme density peaks near the accretors, and (2) the treatment of boundary conditions, where we implement refilling and modified outflow conditions instead of the wave-damping open boundaries used by \citetalias{li22}. Wave-damping boundary conditions rely on fine-tuning parameters, such as the damping timescale, whose selection can significantly affect the outcome of the simulation. Moreover, the effectiveness of wave damping is strongly dependent on the initial conditions; poorly defined or inconsistent initial states can introduce artifacts that the damping scheme may not adequately suppress the perturbations. While wave-damping boundaries can be useful in some scenarios, they may not be suitable for all types of flows, particularly in highly turbulent or chaotic regimes.
\FloatBarrier
\begin{figure}[!ht]
\centering
\includegraphics[width=\columnwidth]{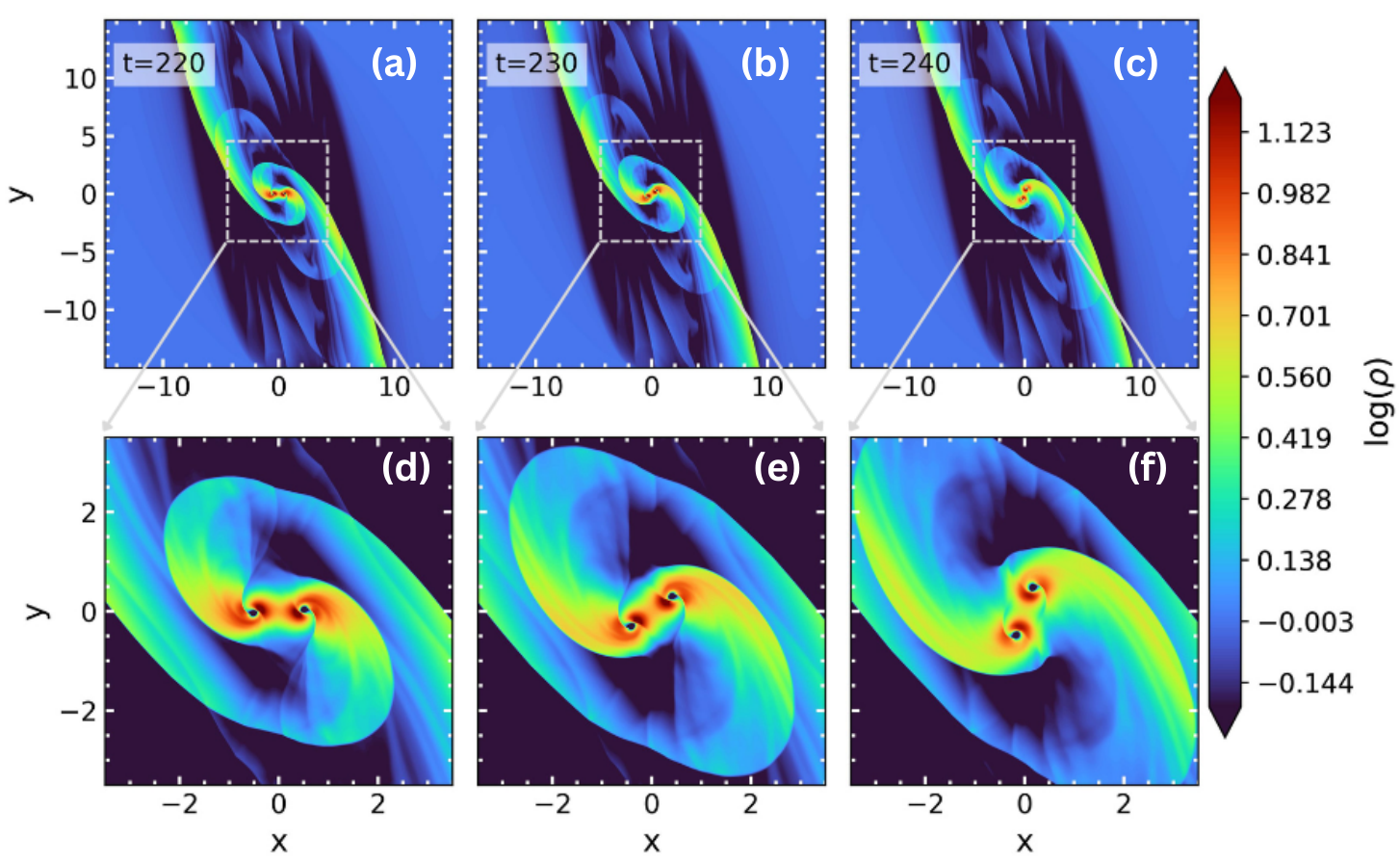}
\caption{Logarithmic density plots for 2D Hydrodynamical run}
\label{2d_hydro}
\end{figure}
\FloatBarrier
Overall, while our simulation and that of \citetalias{li22} employ different codes and boundary conditions, the resulting flow structures are in strong qualitative agreement. In Fig.~\ref{2d_hydro}, the top row shows the evolution of the large-scale disk, where prominent spiral shocks and tidal streams persist as the binary orbits. The bottom row provides zoomed-in snapshots of the central region, clearly illustrating the formation of dense, well-defined CSDs around each black hole and the continuous accretion streams feeding them from the disk. These characteristic features—spiral arms, accretion streams, and CSDs—closely resemble those found by \citetalias{li22}, lending further confidence to the robustness of our numerical setup and chosen parameters. \\
For a more thorough analysis of 2D flow topology, we refer the reader to \cite{Fung_2015} and \cite{Ormel2013}. We must point out that there can be quantitative differences in diagnostic quantities reported by \citetalias{li22}, such as accretion rates, torque, energy transfer rate, and orbital period, due to the different choices of simulation codes and adopted resolution.\

\section{Influence of disk parameters on spiral arms}
\label{sec:app2}

\begin{figure}[!ht]
\centering
\includegraphics[width=\columnwidth]{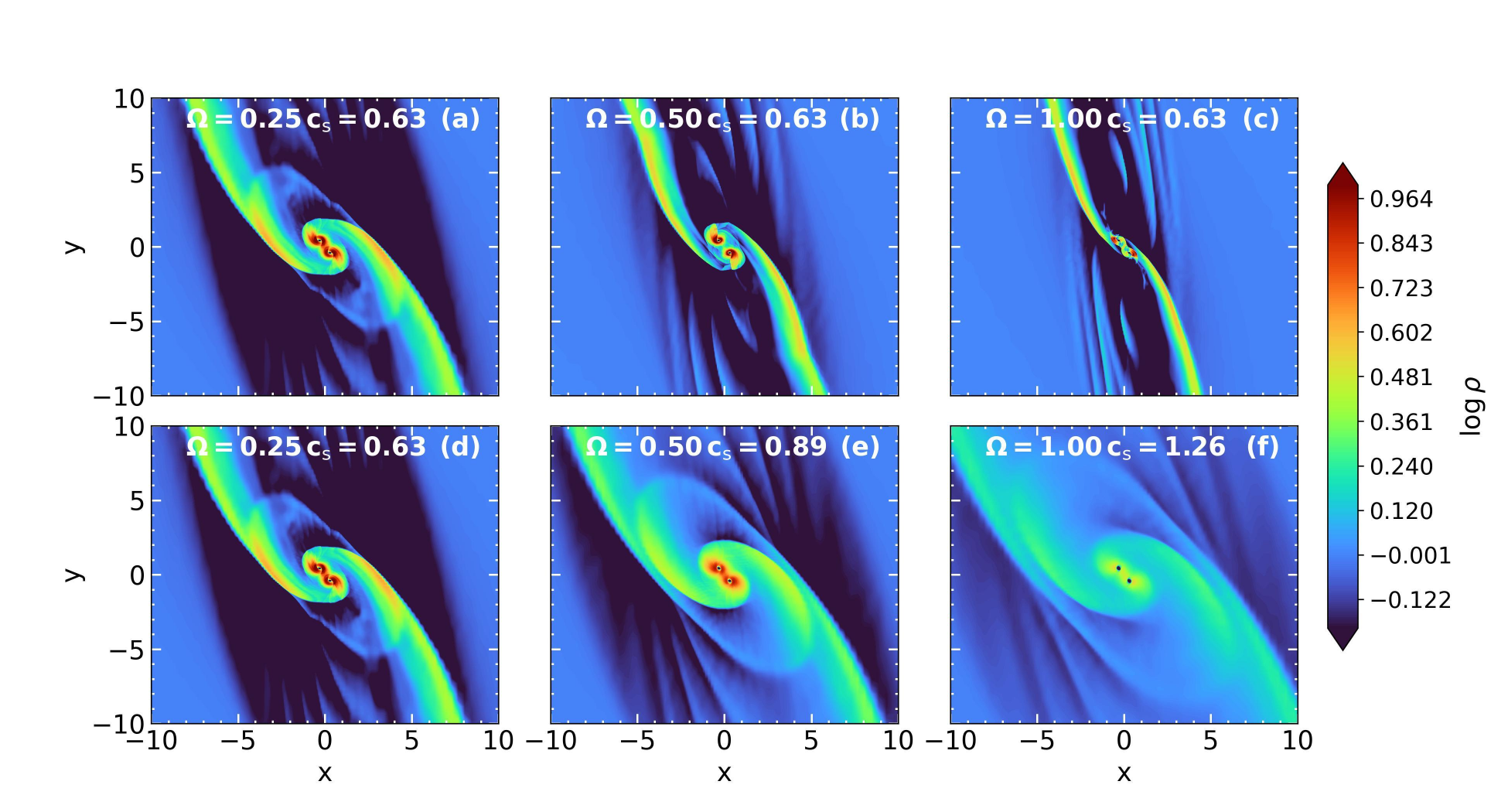}
\caption{Effect of disk parameters on spiral arm morphology.
}
\label{2d_arms}
\end{figure}

In our main simulation suite, both $\Omega_{\text K}$ and $c_\text{s}$ vary simultaneously (see Table \ref{tab:param}), making it difficult to isolate the effect of each parameter. To clarify their individual influence, we performed additional hydrodynamical runs with $\lambda=2.5$ where only one parameter was varied at a time.
Fig. \ref{2d_arms} illustrates the influence of the Keplerian angular velocity ($\Omega_{\text K}$) and sound speed ($c_\text{s}$) on the morphology of spiral arms in the circumbinary disk. In the top row, $c_\text{s}$ is held constant while $\Omega_{\text K}$ increases from left to right. 
\FloatBarrier
\FloatBarrier

As $\Omega_{\text K}$ increases, the spiral arms become progressively more tightly wound, due to the faster orbital motion of the disk material. In the bottom row, $\Omega_{\text K}$ is fixed and $c_\text{s}$ increases from left to right. Higher $c_\text{s}$ leads to broader and more diffuse spiral arms, as increased pressure support smooths density contrasts and weakens the spiral density waves. Together, these panels demonstrate that the winding and sharpness of the spiral arms are primarily controlled by $\Omega_{\text K}$ and $c_\text{s}$, respectively, with their combined effect determining the overall morphology of the circumbinary disk.

\end{appendix}
\end{document}